# Magnon Super-Bloch Oscillations and Switchable Transport in Driven Magnonic Crystals

Xiang Liu[1], Xi-guang Wang[1,*], Zhi-xiong Li[1], Xiufeng Han[2], Guang-hua Guo[1,*]

[1]*School of Physics, Central South University, Changsha, 410083, China*

[2]*Beijing National Laboratory for Condensed Matter Physics, Institute of Physics, Chinese Academy of Sciences, Beijing 100190, China*

*Abstract:* Super-Bloch oscillations (SBOs) offer a powerful route to control wave packet propagation in driven lattices, yet their realization in magnonic systems has remained elusive. Here, we predict SBOs of magnon wave packets in periodic magnonic crystals driven by static and ac gradients of magnetic anisotropy field. A static gradient alone induces magnon Bloch oscillation (BO) with a constant carrier frequency. Remarkably, adding a near-resonant MHz-scale ac driving gives rise to a slow large-amplitude envelope oscillation, the hallmark of SBO. At exact resonance, the ac driving enables phase-controlled directional transport with a carrier-frequency shift up to GHz scale. Consequently, the same structure can be switched among three distinct regimes: conventional BO, SBO, and directed transport. Analytical expressions for the oscillation amplitudes, superperiod, drift velocity, and frequency evolution are validated by micromagnetic simulations. Extending to antiferromagnet magnonic crystals, we further demonstrate that these effects drive opposite motions of two oppositely polarized magnon modes in THz range. Our work establishes magnon SBOs as a versatile platform, opening a new avenue for reconfigurable magnonic devices.

* To whom correspondence should be addressed. E-mail: guogh@mail.csu.edu.cn (G.-h Guo); wangxiguang@csu.edu.cn (X.-g Wang)

*Introduction*—Magnons, being charge-neutral excitations, can propagate without Joule heating and mediate spin current over long distances, making them ideal candidates for wave-based computing architectures and the development of energy-efficient spintronic devices. Controlling the propagation and spectrum of magnons is a central goal of magnonics [1–12]. Magnonic crystals, composed of magnetic periodic structures, provide a solid platform for such control: their periodic band structure reshapes magnon dispersion, opens band gaps, and enables programmable propagation [13-22]. A key phenomenon for wave packet dynamics in periodic media is the Bloch oscillation (BO), which occurs when a static force drives the wave vector across the Brillouin zone. With the Bragg reflection at the Brillouin zone edge, the wave packet does not move constantly but oscillates in space. BOs were originally predicted for electrons [23], and then observed in other (quasi-)particles, including photons in photonic crystals [24], phonons in acoustic lattices [25-27], Cooper pairs in Josephson junctions [28-30], plasmons in graphene [31], and effective anyons in electric-circuit simulators [32]. In magnonic systems, BOs have been studied in graded lattices, nonlinear spin soliton dynamics, and continuous nanostructured media [33–37].

A much richer phenomenon is the super-Bloch oscillation (SBO), where an additional ac drive near the BO frequency produces a slow, large-amplitude envelope oscillation superimposed on the fast BO. SBO has been established in several driven-wave platforms, such as ultracold atomic gases and photonic lattices [38 – 42], showcasing enhanced control over wave packet dynamics. However, its magnonic counterpart has not yet been reported. Utilizing SBO to control the magnon propagation has remained unexplored. In addition, how periodic driving reshapes the carrier frequency of a propagating wave packet is still unclear.

In this letter, we show that a driven magnonic crystal supports three regimes in a single structure: conventional BO under a static gradient, SBO under a detuned ac gradient, and phase-controlled directional propagation at exact resonance. We further show that periodic driving controls not only the packet motion trajectory but also its carrier frequency in time. Analytical solutions for the oscillation amplitudes, superperiod, drift velocity, and frequency evolution are confirmed by micromagnetic simulations. We further apply the mechanism for antiferromagnet, allowing for the polarization-dependent magnon transport within THz range. The predicted effect offers a route toward reconfigurable magnonic devices, such as delay lines, routers, polarizers, and frequency shifters.

*Theoretical model*—The magnonic crystal studied here is a ferromagnetic film with spatially alternating magnetic anisotropy field between $H_{K1}$ and $H_{K2}$ as shown in Fig. 1(a). Similar magnonic crystals have been demonstrated under elastic strain, electric field, ion beam bombardment, or geometry shape [43-46]. The magnon dynamics in the magnonic crystal is governed by the Landau-Lifshitz-Gilbert (LLG) equation,

$$\frac{\partial \mathbf{m}}{\partial t} = -\gamma \mathbf{m} \times \mathbf{H}_{\text{eff}} + \alpha \mathbf{m} \times \frac{\partial \mathbf{m}}{\partial t}, \quad (1)$$

where $\mathbf{m}$ is the unit vector of the local magnetization, and $\alpha$ is the Gilbert damping. The effective magnetic field $\mathbf{H}_{\text{eff}} = (2A_{ex}/\mu_0 M_s)\mathbf{m} + H_A m_x \mathbf{e}_x$ consists of the exchange field with exchange constant $A_{ex}$, and the anisotropy field with spatially and temporally varying coefficient $H_A$. In the numerical simulations, we adopt following parameters of YIG: $M_s = 1.4 \times 10^5$ A/m, $A_{ex} = 3 \times 10^{-12}$ J/m, $H_{K1} = 1 \times 10^5$ A/m, $H_{K2} = 2 \times 10^5$ A/m, and $\alpha = 1 \times 10^{-4}$. The LLG equation is numerically solved employing a fifth-order Runge-Kutta scheme with a time step of 0.01 ps, and the magnonic crystal is discretized with a cell size of 2 nm. Initially, the equilibrium magnetization is in the $+\mathbf{e}_x$ direction. Then, a periodic field pulse $h_a \mathbf{e}_z \sin(2\pi f_H t)/(2\pi f_H t)$ is applied locally at $x = 0$, to excite magnon over a wide frequency range, with an amplitude $h_a = 5000$ A/m and cutoff frequency $f_H = 30$ GHz. Applying Fourier transform on the magnetization $m_z(x, t)$ gives the magnon bands in Fig. 1(b). The periodically varying anisotropy field folds the magnon dispersion at Brillouin zone (BZ) boundaries $k = \pm\pi/a$, where $a$ is the lattice period length. At the BZ boundaries, clear band gaps are generated. With a period of $a = 400$ nm, the lowest magnon band is fitted by $\omega = \omega_0 - 2\beta\cos(ka)$ with width $2\beta = 2\pi \times 0.08$ GHz and central frequency $\omega_0 = 2\pi \times 7.32$ GHz, and higher bands have larger width.

*Results and discussions*—We start with a linearized analytical model for the magnon dynamics under a static anisotropy gradient $H_0x$ and an ac gradient $H_1x\cos(\omega_f t - \varphi)$. By considering small deviations $\delta\mathbf{m} = (0, m_y, m_z)$ around $\mathbf{m}_0 = \mathbf{e}_x$, introducing $\psi = m_y - im_z$ and linearizing the LLG equation, we deduce the Schrödinger equation for magnon,

$$i\frac{\partial\psi}{\partial t} = -\gamma\frac{2A_{\mathrm{ex}}}{\mu_0 M_s}\nabla^2\psi + \left[\gamma H_K(x) + F(t)x\right]\psi \cdot \tag{2}$$

Here, the influence of Gilbert damping is neglected. The effective magnon potential is determined by the anisotropy field. The periodic potential component $\gamma H_K(x)$ brings a series of allowed bands and bandgaps, as demonstrated by Fig. 1(b). To analyze the essential physics, we first project the dynamics onto the lowest magnon band $\omega = \omega_0 - 2\beta\cos(ka)$. According to the acceleration theorem $dk(t)/dt = -F_0$ [47], under the static gradient $F = F_0 = \gamma H_0$, the wavevector evolution follows $k(t) = k_0 - F_0 t$, where $k_0$ is the initial magnon wavevector at time $t = 0$. Together with the magnon group velocity $v_g = d\omega/dk = 2\beta a\sin[k(t)a]$, the propagating magnon wave packet center $x(t)$ is derived as,

$$x(t) = x_0 + \int_0^t v_g(t')dt' = x_0 + \frac{2\beta a}{\omega_B}\left[\cos(\omega_B t - k_0 a) - \cos(k_0 a)\right] \cdot \tag{3}$$

Here, $x_0$ is the initial magnon position at $t = 0$. The magnon wave packet center executes BO with frequency $\omega_B = F_0 a$ and amplitude $2\beta / F_0$. The packet carrier frequency remains constant during BOs as $\frac{\partial\omega}{\partial t} = \frac{\partial\omega}{\partial(F_0 x)}\frac{\partial(F_0 x)}{\partial t} + \frac{\partial\omega}{\partial k}\frac{dk}{dt} = F_0 v_g - F_0 v_g = 0$ with the magnon dispersion $\omega(k) = \omega_0 + F_0 x - 2\beta\cos(ka)$ in the gradient. In such scenario, the change in potential energy is exactly compensated by the change in band energy.

Micromagnetic simulations based on the LLG Eq. (1) confirm such effect. At $t = 0$, we initially inject a Gaussian pulse magnon packet propagating in the $-x$ direction centered at $x_0 = 0$, with full width at half maximum of 4 μm and frequency at $\omega_0 = 2\pi \times 7.32$ GHz. Without the static gradient, the frequency is at the center of magnon band with $k_0 a \approx -\pi/2$, and the magnon packet can freely propagate in the magnonic crystal. Under the gradient $H_0 = 0.7$ [(A/m)/nm], the wave packet undergoes BO with frequency $\omega_B = F_0 a = 2\pi \times 0.01$ GHz and amplitude of about $2\beta/F_0 = 4$ μm, in good agreement with Eq. (3) [Fig. 1(c)]. The small deviation from the analytic curve can be attributed to the finite packet width and to the slight departure of the lowest band from an ideal cosine dispersion.

When a near resonant ac gradient $F_1\cos(\omega_f t - \varphi)$ (with $\omega_f$ close to $\omega_B$) is added, the SBO is generated. The acceleration theorem gives $k(t) = k_0 - F_0 t - (F_1/\omega_f)\sin(\omega_f t - \varphi)$. With Jacobi-Anger expansion, harmonics $\omega_B + n\omega_f$ appear in the group velocity $v_g = 2\beta a\sin[k(t)a]$, where the integer $n = 0, \pm1\ldots$. As the detuning $\delta = \omega_B - \omega_f << \omega_B$ is small, the $n = -1$ term gives rise to a slow envelope with amplitude $A_1 = 2(\beta a/\delta)J_1(F_1 a/\omega_f)$ and superperiod $T_{\text{super}} = 2\pi/|\delta|$, on top of the fast BO with amplitude $A_0 = 2(\beta a/\omega_B)J_0(F_1 a/\omega_f)$, yielding SBO in the form of

$$x(t) = x_0 + A_0\left[\cos(\omega_B t - k_0 a) - \cos(k_0 a)\right] - A_1\left[\cos(\delta t + \varphi - k_0 a) - \cos(\varphi - k_0 a)\right]. \quad (4)$$

Here, $J_{1(0)}(F_1 a/\omega_f)$ is the Bessel function of the first kind with integer order 1(0). Without the ac force $F_1$, there are $J_0(F_1 a/\omega_f) = 1$ and $J_1(F_1 a/\omega_f) = 0$, and Eq. (4) returns back to Eq. (3) for BO. In contrast to BO, the packet carrier frequency becomes time dependent in SBO regime,

$$\frac{\partial\omega}{\partial t} = \frac{\partial\omega}{\partial(Fx)}\frac{\partial(Fx)}{\partial t} + \frac{\partial\omega}{\partial k}\frac{dk}{dt} = -F_1\omega_f \sin\left(\omega_f t - \varphi\right)x(t). \quad (5)$$

Thus, the driven magnonic crystal controls both the real space motion and the spectral evolution of the propagating magnon packet.

The driven SBOs are verified quantitatively in the simulations. For $H_0 = 0.7$ [(A/m)/nm], $H_1 = 0.2$ [(A/m)/nm] and $\omega_f = 2\pi \times 0.012$ GHz in Figs. 2(a-b), the packet exhibits SBO with $|A_0| \approx 3.9$ μm, $|A_1| \approx 2.3$ μm and $T_{\text{super}} \approx 500$ ns, together with a pronounced oscillation of the carrier frequency. Increasing the drive to $H_0 = 1.5$ [(A/m)/nm] ($\omega_B = 2\pi \times 0.021$ GHz), $H_1 = 0.8$ [(A/m)/nm] and $\omega_f = 2\pi \times 0.019$ GHz in Figs. 2(c-d), the fast BO component is suppressed to $|A_0| \approx 1.7$ μm and the slow SBO envelope is enhanced to $|A_1| \approx 5.4$ μm. We further analyze the influences of parameters $H_0$, $H_1$ and $\delta$ following Eq. (4). The results are shown in Figs. 2(e-g). Through the Bessel factor, increasing $H_1$ weakly reduces $|A_0|$ but enhances $|A_1|$. Increasing the static force under a fixed $\delta = \omega_B - \omega_f = -2\pi \times 0.002$ GHz, reduces both $|A_0|$ and $|A_1|$. With increasing $|\delta|$, $|A_0|$ remains unchanged while $|A_1|$ is suppressed.

At exact resonance $\delta = 0$, the detuned envelope breaks down and is replaced by resonant directional propagation. In this case, the magnon packet motion becomes,

$$x(t) = x_0 + A_0\left[\cos\left(\omega_B t - k_0 a\right) - \cos\left(k_0 a\right)\right] + 2\beta a J_1\left(\frac{F_1 a}{\omega_f}\right)\sin\left(\varphi - k_0 a\right)t, \tag{6}$$

which shows a drift of magnons with velocity $v_m = 2\beta a J_1(F_1 a/\omega_f)\sin(\varphi - k_0 a)$, accompanied by BO. The direction and magnitude of $v_m$ are set by the phase $\varphi$ of the ac drive and initial wavevector $k_0$. For the injected magnon packet with $k_0 a \approx -\pi/2$, the phase control maps onto the forward motion at $\varphi = 0$, near trapping at $\varphi = -\pi/2$, and reverse motion at $\varphi = -\pi$, as observed in Figs. 3(a, c, e). For $\varphi = -\pi/2$ in Fig. 3(c), the difference between simulation and calculation is ascribed to the band curve difference from $\omega_0 - 2\beta\cos(ka)$, and to the fact that the simulated $k_0 a$ deviates from $-\pi/2$ at 7.32 GHz. If we set $k_0 a = -0.35\pi$, a better agreement is achieved for the green curve. Figure 3(g) further demonstrates the phase dependence of the averaged drift velocity, with a tunable range of approximately ±50 m/s in the simulated parameter range. As the drift velocity is directly proportional to the bandwidth $\beta$, faster motion is achievable for wider band.

The carrier frequency also varies under resonant driving. Here, the time-dependent magnon frequency is solved as,

$$\omega = \omega_0 + C_0 t + \frac{(C_1 - C_2 t)}{\omega_B}\cos\left(\omega_B t - \varphi\right) - \frac{C_2}{\omega_B^2}\sin\left(\omega_B t - \varphi\right), \tag{7}$$

where the constants $C_0 = F_1 A_0 \omega_B \sin(\varphi - k_0 a)/2$, $C_1 = F_1 \omega_f x_0 + C_0\cos(k_0 a)$, and $C_2 = 2F_1\omega_f \beta a J_1(F_1 a/\omega_f)\sin(\varphi - k_0 a)$. Equation (7) predicts a phase-dependent frequency shift ($C_0$ term) with an oscillatory modulation ($C_1$ and $C_2$ terms) superimposed on it, in agreement with the time-windowed spectra in Figs. 3(b, d, f). Meanwhile, the $C_2 t$ term brings gradually increasing amplitude in the frequency fluctuation. In the simulated parameter range, the resulting frequency steering reaches the GHz scale even though the applied drive frequency is only ~0.02 GHz. Exact resonance driving therefore converts a low-frequency control field into strong, phase-selective steering of both packet motion and carrier frequency. Furthermore, we prove that the motion and carrier frequency variation trajectories are dynamically reconfigurable by changing the ac drive phase $\varphi$, as detailed in section I in the supplementary material (SM).

The BO, SBO and directional propagation mechanisms are operational in antiferromagnet (AFM) magnonic crystal, enabling selective manipulation of the two left- and right-hand circularly polarized magnon modes in the THz range. For the AFM magnonic crystal detailed in section II in the SM the magnon bands for two opposite polarization modes are degenerate as shown in Fig. 4(a). Including static and ac gradient magnetic fields, we find that the BOs, directional propagations and SBOs of the left- and right-hand magnons occur in opposite directions, as demonstrated in Fig. 4(b-d). Due to the narrow band gap between the lower and upper bands, the excited magnon packet in the lower band is accompanied by a weaker excitation in the upper band, which slightly affects the simulation results in Fig. 4(b-d). The opposite behavior of the two magnon modes originates from spatial inversion symmetry breaking induced by the magnetic field. In contrast, when we apply a gradient electric field or strain to drive a gradient anisotropy field, the motions of left-hand and right-hand magnons become identical (not shown). This polarization-dependent response is crucial for achieving complete separation of the two magnon modes.

Concerning the intrinsic Gilbert damping induced magnon attenuation during the manipulation, we suggest one can implement the anti-damping spin torque via electric current pulse, to amplify the attenuated magnon for convenience of observation. Besides, the above effects are further applied to dipolar interaction dominated magnon mode, which are shown in sections III and IV in the SM. Adopting a larger bandwidth $\beta$, BO and SBO amplitudes in the millimeter range are achieved, and the maximum net velocity in the directional motion can reach 3 km/s. The change of the BO and SBO amplitudes and drift velocity of magnon packet from narrow bandwidths to large bandwidths validates the mechanism's scalability from the micrometer to the millimeter range.

*Conclusions*—In summary, we predict magnon SBOs in driven ferromagnetic and antiferromagnetic magnonic crystals, where periodic driving provides simultaneous control over packet trajectory and carrier frequency. A static gradient induces magnon BOs with fixed carrier frequency, a near resonant ac gradient gives rise to large-amplitude SBOs, and an ac gradient at resonance enables phase sensitive directional transport. Analytical expressions for the superperiod, oscillation amplitudes, drift velocity, and frequency evolution are verified by micromagnetic simulations. In AFM magnonic crystals, the magnon BO, SBO and directional transport become polarization dependent, enabling efficient separation and independent control of oppositely polarized magnon modes. As the required gradients can be implemented through magnetic field, strain, or voltage control of anisotropy, the mechanism offers a route toward reconfigurable magnonic delay lines, routers, and frequency shifters.

*Acknowledgment*—This work was supported by the National Natural Science Foundation of China (No. T2495212, No. 12274469, No. 12174452, and No. 12074437), the Natural Science Foundation of Hunan Province of China (Grants No. 2025JJ20005), and the Central South University Innovation-Driven Research Program (Grant No. 2023CXQD036). We are grateful for resources from the High-Performance Computing Center of Central South University.

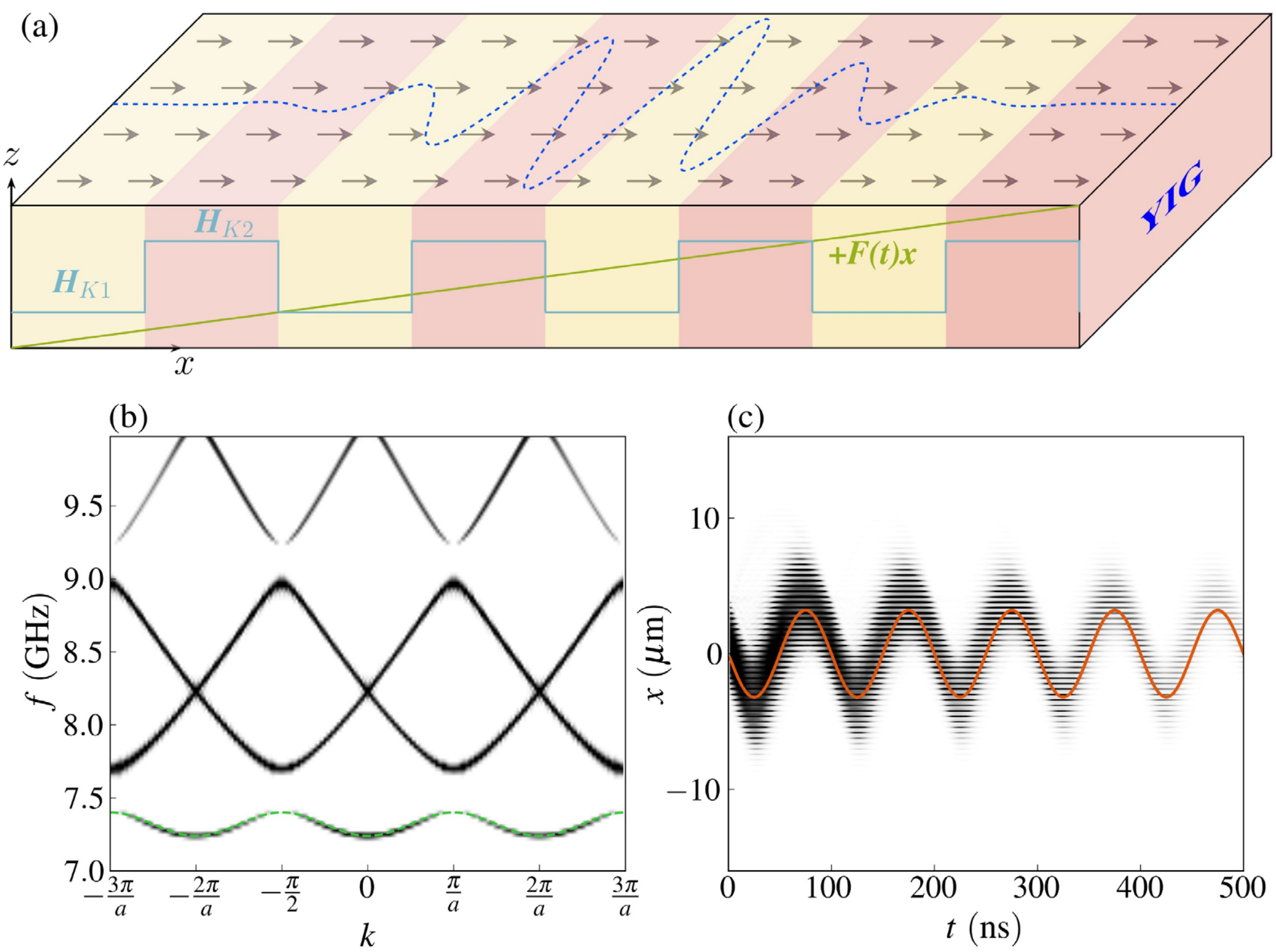

FIG. 1. (a) Schematic of the magnonic crystal with spatially periodic varying anisotropy field. The Gaussian pulse magnon packet is injected at the center $x = 0$, experiencing the BO and SBO under gradient field. (b) Simulated magnonic dispersion relation. The bottom band is fitted by the green dashed line of $\omega = \omega_0 - 2\beta\cos(ka)$ with $2\beta = 2\pi \times 0.08$ GHz and $\omega_0 = 2\pi \times 7.32$ GHz. (c) Motion trajectory of magnon packet obtained from numerical simulation when the static gradient $H_0 = 0.7$ [(A/m)/nm]. The black color represents the simulated results and the orange solid line is calculated from Eq. (3).

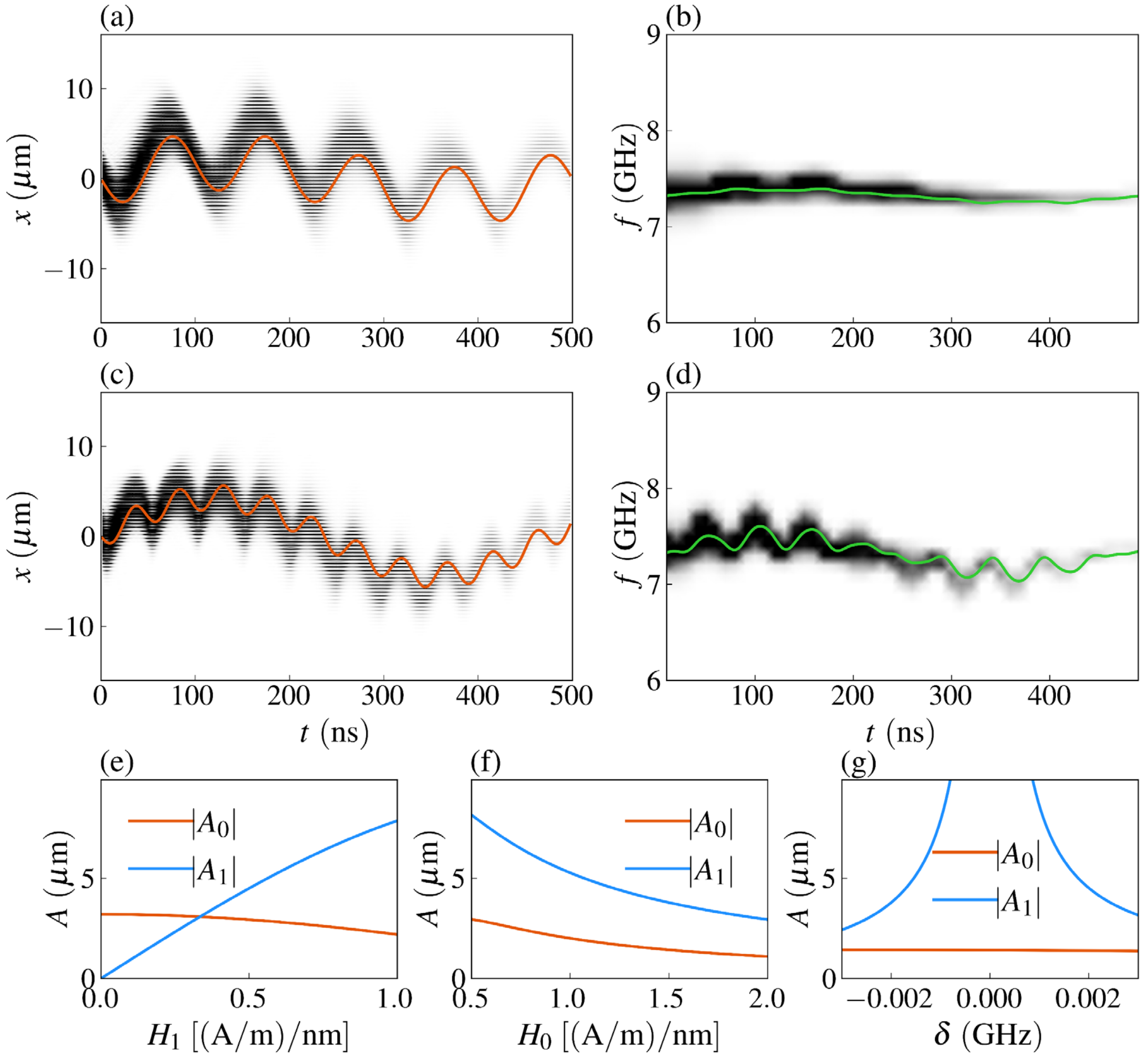

FIG. 2. Motion trajectories and time-windowed frequency spectra of the magnon packet when (a-b) $H_0$ = 0.7 [(A/m)/nm], $H_1$ = 0.2 [(A/m)/nm] and $\omega_f = 2\pi \times 0.012$ GHz, and when (c-d) $H_0$ = 1.5 [(A/m)/nm], $H_1$ = 0.8 [(A/m)/nm] and $\omega_f = 2\pi \times 0.019$ GHz ($\delta = \omega_B - \omega_f = 2\pi \times 0.002$ GHz). The time-windowed frequency spectrum is calculated from the Fourier transformation of the magnetization oscillation in every 10 ns. The orange and green solid lines are calculated from Eqs. (4) and (5). (e-g) Amplitudes $|A_0|$ and $|A_1|$ as functions of $H_1$ with $H_0$ = 0.7 [(A/m)/nm] and $\delta = -2\pi \times 0.002$ GHz, as functions of $H_0$ with $H_1$ = 0.8 [(A/m)/nm] and $\delta = -2\pi \times 0.002$ GHz, and as functions of $\delta$ with $H_0$ = 1.5 [(A/m)/nm] and $H_1$ = 0.8 [(A/m)/nm], respectively.

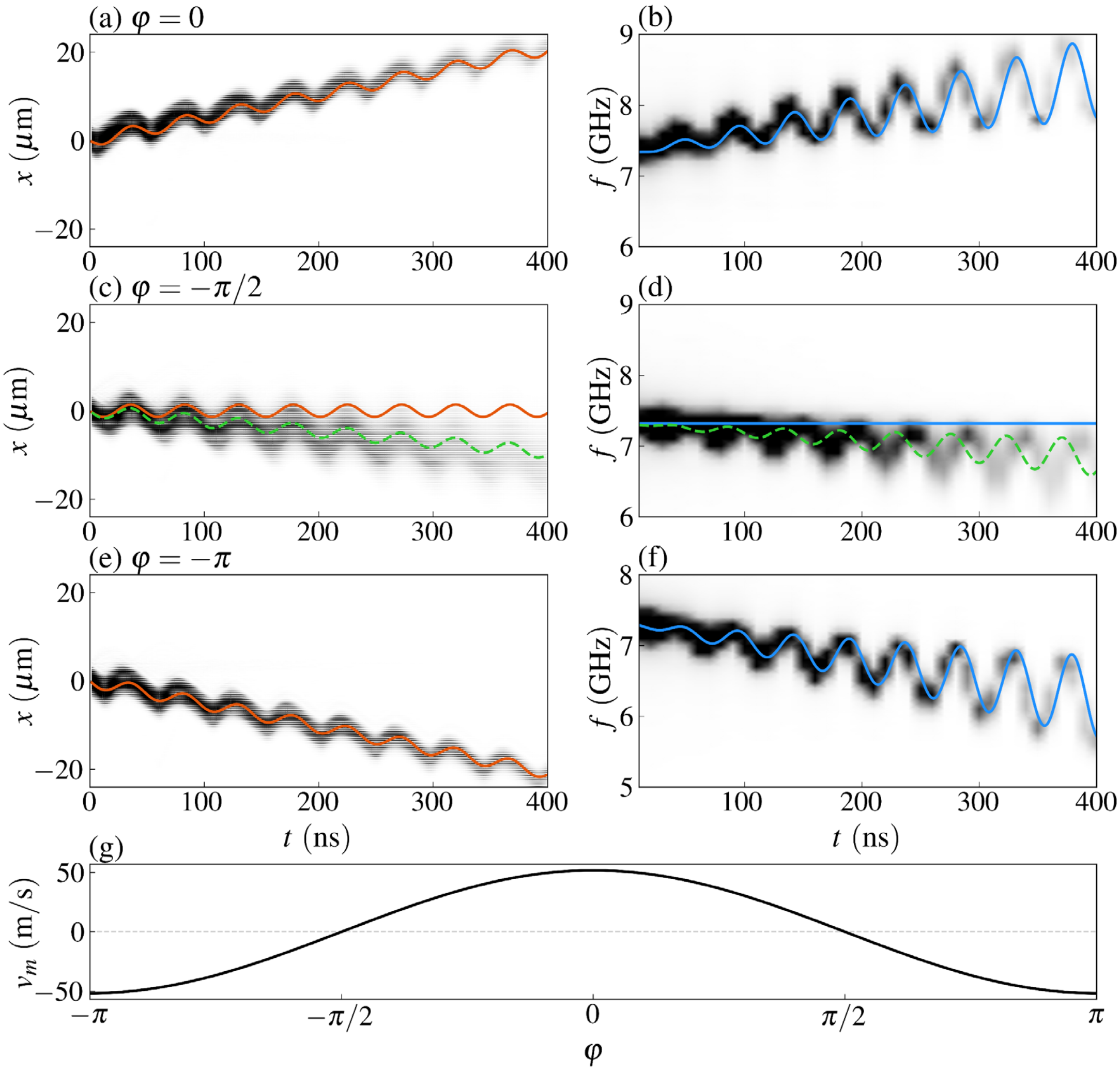

FIG. 3. Motion trajectories and frequency spectra of the magnon packet for (a-b) $\varphi = 0$, (c-d) $\varphi = -\pi/2$, and (e-f) $\varphi = -\pi$ when $H_0 = 1.5$ [(A/m)/nm], $H_1 = 0.8$ [(A/m)/nm] and $\omega_f = 2\pi \times 0.021$ GHz ($\delta = 0$). The orange and blue solid lines are calculated from Eqs. (6) and (7), and the green dashed lines are for $\varphi = -0.35\pi$. (g) Average magnon velocity $v_m$ as a function of initial phase $\varphi$.

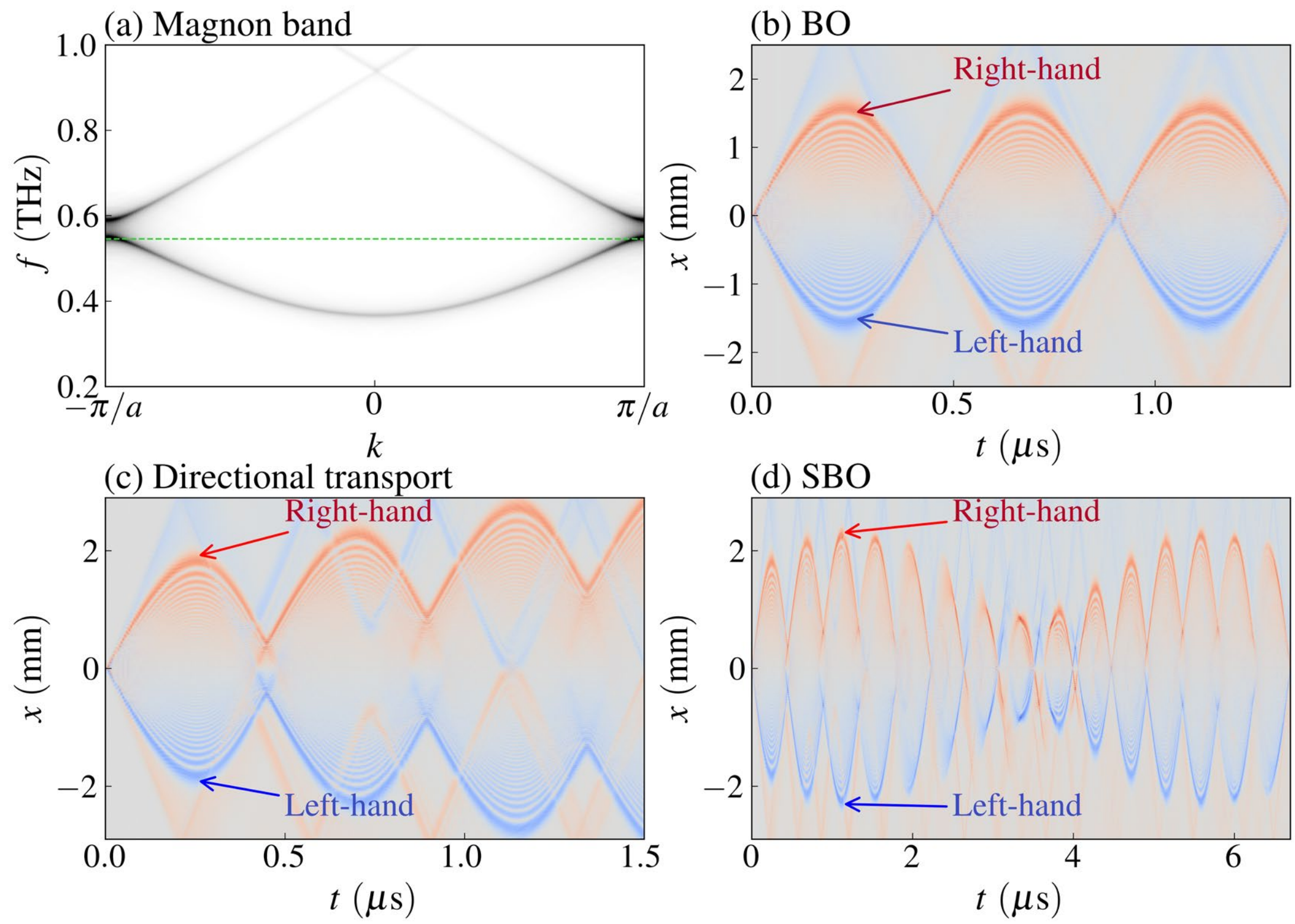

FIG. 4. (a) Band structure for the AFM magnonic crystal. (b) Magnon BO under the static magnetic field gradient 0.004 T/nm with $\omega_B = 2.2$ MHz. (c) Magnon directional transport driven by the ac gradient 0.0008 T/nm with $\omega_f = \omega_B$ and initial phase $\varphi = -\pi/2$. (d) Magnon SBO driven by the ac gradient 0.0008 T/nm with $\omega_f - \omega_B = 0.1\omega_B$. Initially, the magnon packet with frequency 0.55 THz is injected at $x = 0$. The red and blue colors represent the simulated amplitudes of the right-hand and left-hand magnon modes, respectively.